\documentclass{aastex}

\RequirePackage{color}

\begin{document}

\title{The Localized Energy Distribution of Dark Energy Star Solutions}
\shorttitle{Dark Energy Stars}
\shortauthors{Halpern et al.}

\author{Paul Halpern} \and \author{Michael Pecorino}
\affil{Department of Mathematics, Physics and Statistics,\\
University of the Sciences in Philadelphia,\\600 S. 43rd St.,\\
Philadelphia, PA. 19104, USA}

\begin{abstract}
We examine the question of energy localization for an exact solution of Einstein's equations with a scalar field corresponding to the phantom energy interpretation of dark energy.  We apply three different energy-momentum complexes, the Einstein, Papapetrou and M{\o}ller prescriptions, to the exterior metric and determine the energy distribution for each.  Comparing the results, we find that the three prescriptions yield identical energy distributions.

\vspace{.5 cm}

Published in ISRN Astronomy and Astrophysics, Vol. 2013.
\end{abstract}

\keywords{black hole; phantom energy; Schwarzschild solution; dark energy; energy-momentum complex}

\section{Introduction}

The question of energy localization has been a pressing issue in general relativity since Einstein formulated his field equations.  Einstein sought to include the energy and
momenta of gravitational fields along with those of matter and
non-gravitational fields as a well-defined locally conserved
quantity.  To that end, he proposed an energy-momentum complex that follow conservation laws for those quantities \citep{einstein}.

One concern about the Einstein energy-momentum is that its value depends on the coordinate system used.  Specifically, it favors the use of quasi-Cartesian (perturbations of a flat background) coordinates.  Accordingly, it does not transform as a tensor.  Because it is asymmetric in its indices, it does not conserve angular momentum.

To address these issues, a number of other theorists have proposed alternative definitions for energy-momentum complexes.  These include prescriptions by Papapetrou \citep{papapetrou}, Landau and Lifshitz \citep{landau}, Weinberg \citep{weinberg} and others, each designed to conserve angular momentum in addition to energy-momentum.  A prescription by M{\o}ller \citep{moller1, moller2} offers the special advantage of being coordinate-system independent.

While at first the multiplicity of energy-momentum complexes discouraged theorists from using them, important results in the 1990s suggested an underlying unity that served to revive interest.  In 1990,  Virbhadra found consistency in the application of distinct energy-momentum prescriptions to the same metric \citep{virbhadra1}.  Six years later, a paper by Aguirregabiria, Chamorro and Virbhadra revealed that all metrics
of the Kerr-Schild class, including the Schwarzchild, Reissner-Nordstr{\"o}m and other
 solutions, yield identical energy and momentum distributions for a variety of prescriptions  \citep{acv}.  Then in 1999, a paper by Virbhadra indicated the energy-momentum complexes of Einstein, Papapetrou, Weinberg, and Landau and Lifshitz produced similar results for metrics more general than the Kerr-Schild class if they are constructed with Kerr-Schild Cartesian coordinates \citep{virbhadra2}.  

There have subsequently been many attempts to apply the various energy-momentum complexes to black holes and related astronomical objects.  In recent years, for example, Radinschi, along with various colleagues, has investigated the energy-momentum distribution for stringy black holes \citep{radinschi6, radinschi2}, Ho{\^r}ava-Lifshitz black holes \citep{radinschi4} , for charged black holes in generalized dilaton-axion gravity  \citep{radinschi5}, for asymptotically de Sitter spacetimes  \citep{radinschi1}, and in a Schwarzschild-quintessence spacetime  \citep{radinschi3}.  Vagenas has examined the energy distribution in the dyadosphere of a Reissner-Nordstr{\"o}m black hole \citep{vagenas}.  Berman has looked at the energy and angular momentum of dilation black holes \citep{berman}.  Sharif has examined the energy of rotating spacetimes in teleparallel gravity \citep{sharif}.  Halpern has investigated the energy distribution of a charged black hole with a minimally coupled scalar field \citep{halpern}.

In this paper, we consider yet another scenario:  a static, electrically-neutral, spherically-symmetric massive object (such as a star or black hole) embedded in a space filled with phantom energy.  Phantom energy is an especially potent type of dark energy proposed by Caldwell \citep{caldwell} that possesses an equation of state parameter $w < -1$.  It and other forms of dark energy have been hypothesized as possible agents for the acceleration of the universe, as recorded in supernova surveys \citep{riess, perlmutter}.

The interior and exterior metrics of a dark energy star were calculated by Yazadjiev \citep{Yazadjiev}.  The metrics depend on a mass parameter $m$, the radius $r$ and an additional parameter $\beta$.  From these, Yazadjiev determined an overall mass $M$ (that includes the rest mass as well as dark energy) and a dark charge $D$, through the following relationships:

\begin{eqnarray}
M &=& \cosh{\beta}  \hspace{0.2 cm} m \\
D &=& \sinh{\beta} \hspace{0.2 cm} m
\end{eqnarray}

Note that $\beta = 0$ corresponds to the situation of a conventional Schwarzschild solution without phantom energy.  The conventions that $c = 1$ and $G = 1$ are followed.

Applying the prescriptions of Einstein, Papapetrou and M{\o}ller to the exterior metric derived by Yazadjiev we will examine the localized energy of the dark energy star.

 \section{Applying Einstein's prescription to dark energy stars}

We now apply Einstein's prescription to Yazadjiev's dark energy star solution.  The exterior metric has the following form:

 \begin{eqnarray}
ds^2 &=& - {(1 - 2 m/r)}^\kappa dt^2 \nonumber \\
 & & + {(1 - 2 m/r)}^{1-\kappa} [\frac{dr^2}{1-\frac{2m}{r}} \nonumber \\ & & +r^2({d\theta}^2 + {\sin{\theta}}^2 d{\phi}^2)]
 \end{eqnarray}
 
 where:
 \begin{equation}
 \kappa = \cosh{\beta} = \frac{M}{\sqrt{M^2-D^2}}
 \end{equation}

Einstein's prescription defines the local energy-momentum density
as:

\begin{equation}
{{\theta}_i}^k = \frac{1}{16 \pi} {{H_i}^{kl}}_{,l}
\end{equation}

where the superpotentials ${H_i}^{kl}$ are given by:

\begin{equation}
{H_i}^{kl} = \frac{g_{in}}{\sqrt{-g}}{[-g (g^{kn} g^{lm} - g^{ln} g^{km})]}_{,m}
\end{equation}

This complex has the antisymmetric property that:
\begin{equation}
{H_i}^{kl} = -{H_i}^{lk}
\end{equation}

The energy-momentum components can be found by integrating the
energy-momentum density over the volume under consideration:
\begin{equation}
P_i = \int \int \int {\theta_i}^0 dx^1 dx^2 dx^3
\end{equation}

Through Gauss's theorem we can express this as a surface integral:

\begin{equation}
P_i = \frac{1}{16 \pi} \int \int {H_i}^{0 \alpha} {\mu}_{\alpha} dS
\end{equation}

where ${\mu}_{\alpha}$ is the outward unit vector normal to the
spherical surface element:
\begin{equation}
dS = r^2 \hspace{0.1 cm}\rm{sin}\theta \hspace{0.1 cm} d\theta
\hspace{0.1 cm} d\phi
\end{equation}

The localized energy $E = P_0$ can thereby be expressed as:

\begin{equation}
P_0 = \frac{1}{16 \pi} \int \int {H_0}^{0 \alpha} {\mu}_{\alpha} dS
\end{equation}

Therefore, the relevant superpotentials to determine the localized energy are ${H_0}^{0 \alpha}$ with $\alpha$ ranging from $1$ to $3$. 

Taking $i= k = 0$ we find that (6) reduces to:

\begin{equation}
{H_0}^{0 l} = \frac{g_{0n}}{\sqrt{-g}}{[-g (g^{0n} g^{lm} - g^{ln} g^{0m})]}_{,m}
\end{equation}

Note that $g_{0n} = 0$ for all values of $n$ except $n = 0$  Therefore, the only non-zero components of (12) are the ones for which $n = 0$, namely:

\begin{equation}
{H_0}^{0l} = \frac{g_{00}}{\sqrt{-g}}{[-g (g^{00} g^{lm} - g^{l0} g^{0m})]}_{,m}
\end{equation}

We substitute the metric (3) into expression (13) and find that the relevant superpotential values are:

\begin{eqnarray}
{H_0}^{01} &=&  \frac{4 \kappa m x}{r^3}\\
{H_0}^{02}&=&  \frac{4 \kappa m y}{r^3} \\
{H_0}^{03} &=&  \frac{4 \kappa m z}{r^3}
\end{eqnarray}

We insert the super potentials (14-16) into the double integral (11).   Evaluating this double integral over the full coordinate range, we find the
total energy of s solution within a sphere of radius $r$  to be:

\begin{equation}
E= P_0 = \kappa m = M
\end{equation}

It is interesting that the localized energy precisely matches the dark energy star's total mass $M$.  This indicates that the Einstein complex well-encompasses the complete energy of the star, including the energy associated with its rest mass $m$ as well as its dark energy.  Note that if the dark charge $D$ is zero, $\kappa$ becomes $1$ and the localized energy reverts to the Schwarzschild value of $m$.

\section{Determining the energy by use of Papapetrou's prescription}

We now turn to a second method for determining the local energy,  Papapetrou's prescription.
Unlike Einstein's prescription,  Papapetrou's energy-momentum
complex offers the advantage of being symmetric in its
indices.  Hence it permits the precise definition of  local conservation laws.

The Papapetrou energy-momentum complex is defined as:

\begin{equation}
{\Omega}^{ik} = \frac{1}{16 \pi} {{N}^{ikab}}_{,ab}
\end{equation}

\begin{equation}
P_i = \int \int \int {\Omega}^{i0} dx^1 dx^2 dx^3
\end{equation}

where the functions ${N}^{ikab}$ are given by:

\begin{equation}
{N}^{ikab} = \sqrt{-g}\hspace{0.1 cm}[g^{ik} {\eta}^{ab} - g^{ia} {\eta}^{kb} +
g^{ab} {\eta}^{ik} - g^{kb} {\eta}^{ia}]
\end{equation}

\vspace{0.5 cm}

and the $\eta^{ab}$ terms represent the components
of a Minkowski metric of signature $-2$.

Again, we use Gauss's theorem to express the total energy as a surface integral:

\begin{equation}
E = \frac{1}{16 \pi} \int \int  {\chi}^{00 \alpha}  {\mu}_{\alpha} dS
\end{equation}

with the superpotentials ${\chi}^{00 \alpha}$ defined as:

\begin{equation}
{\chi}^{00 \alpha} = N^{00k \alpha}_{,k}
\end{equation}

We determine the relevant values of the superpotentials to be:

\begin{eqnarray}
{\chi}^{001} &=& \frac{4 \kappa m x}{r^3} \\
{\chi}^{002}&=& \frac{4 \kappa m y}{r^3}\\
{\chi}^{003} &=& \frac{4 \kappa m z}{r^3}
\end{eqnarray}

Substituting equations (20-22) into (18) yields:

\begin{equation}
E= P_0 = \kappa m = M
\end{equation}

This is identical to the expression obtained using Einstein's prescription.  It is instructive to see that both complexes produce the same result.

 \section{Determining the energy by use of the M{\o}ller prescription}

We now turn to a third method for determining the local energy, the M{\o}ller prescription, which has the marked advantage of being coordinate-system-independent.  The complex defined by M{\o}ller can be expressed as:

\begin{equation}
{\Xi}_i^k = \frac{1}{8 \pi} {\chi}_{i, p}^{kp}
\end{equation}

where:

\begin{equation}
{\chi}_i^{kl} = \sqrt{-g} \hspace{0.1 cm} [g_{ip, q} - g_{iq, p}] \hspace{0.1 cm}
g^{kq} \hspace{0.1 cm} g^{lp}
\end{equation}

Inserting the metric components for  Yazadjiev's solution (3) we obtain:

\begin{eqnarray}
{\chi_0}^{01} &=& \frac{4 \kappa m x}{r^3} \\
{\chi_0}^{02}&=& \frac{4 \kappa m y}{r^3}\\
{\chi_0}^{03} &=& \frac{4 \kappa m z}{r^3}
\end{eqnarray}

One more time, we employ Gauss's theorem to express the total energy as a surface integral:

\begin{equation}
E = \frac{1}{16 \pi} \int \int  {\chi_0}^{0 \alpha}  {\mu}_{\alpha} dS
\end{equation}

Integrating over the full range of coordinates, we find the total energy using the M{\o}ller complex to be:

 \begin{equation}
E=  \kappa m = M
\end{equation}

This is identical to the expression obtained using Einstein's and Papapetrou's prescriptions. 

 \section{Conclusion}

We have determined the localized energy distribution for Yazadjiev's solution representing a static, electrically-neutral, spherically-symmetric massive object with phantom energy.   In applying the Einstein, Papapetrou and M{\o}ller energy-momentum complexes to Yazadjiev's metric, we have found that each yields an identical localized energy equal to the mass $M$.  This extends earlier results for Kerr-Schild objects such as the Schwarzschild solution to an interesting case that could bear upon the dark energy question.  Our findings help broaden the applicability of energy-momentum complexes, augmenting their use in general relativity as consistent, well-defined ways of describing the local distribution of energy.

\acknowledgments

Thanks to K. S. Virbhadra for his helpful advice throughout the years about energy localization and other aspects of general relativity.


\begin{thebibliography} {}
    \bibitem[Aguirregabiria, et. al.(1996]{acv} Aguirregabiria, J. M., Chamorro, A. and K. Virbhadra, K.S. (1996)  Gen. Rel. Grav. 28, 1393.
    \bibitem[Berman(2008)]{berman} Berman, M.S. (2008) Revista Mex. Astron. Astrof. 44, 285.
    \bibitem[(2002)]{caldwell} Caldwell, R. (2002) Phys. Lett. B 545, 23.
    \bibitem[Einstein(1915)]{einstein} Einstein, A. (1915) Preuss. Akad. Wiss. Berlin 47, 778.
    \bibitem[Halpern(2008)]{halpern} Halpern, P. (2008) Astrophys. and Space Sci. 313, 4, 357.
    \bibitem[(1962)]{landau} Landau, L.D. and Lifshitz, E.M. (1962) The Classical Theory of Fields, Pergamon Press, Oxford, 341.
    \bibitem[(1958)]{moller1} M$\phi$ller, C. (1958) Ann. Phys. 4, 347.
   \bibitem[(1961)]{moller2} M$\phi$ller, C. (1961) Ann. Phys. 12, 118.
     \bibitem[(1948)]{papapetrou} Papapetrou, A. (1948) Proc. R. Ir. Acad. A 52, 11.
\bibitem[Perlmutter, et. al.(1999)]{perlmutter} Perlmutter, S. et al. [Supernova Cosmology Project Collaboration] (1999) Astrophys. J. 517,
565.

     \bibitem[Radinschi(2011)]{radinschi1} Radinschi, I. (2011) Cent. Eur. Jour. of Phys 9, 5, 1173.
  \bibitem[Radinschi and Ciobanu(2006)]{radinschi2} Radinschi, I. and Ciobanu, B. (2006) ``Weinberg
    Energy-Momentum Complex for a Stringy Black Hole Solution,''
    gr-qc/0608029.
     \bibitem[Radinschi, et. al.(2012)]{radinschi3} Radinschi, I., Grammenos, T. and Spanou, A.( 2012) ``Distribution of Energy-Momentum in a Schwarzschild-Quintessence Space-time Geometry,'' http://arxiv.org/abs/1204.1663.   
    \bibitem[Radinschi, et. al.(2011)]{radinschi4} Radinschi, I., Rahaman F. and Banerjee, A. (2011) Int. J. Theor. Phys. 50, 9, 2906.
     \bibitem[Radinschi, et. al.(2010)]{radinschi5} Radinschi, I.,  Rahaman, F. and Ghosh, A. (2010) Int. J. Theor. Phys 49, 943.
    \bibitem[Radinschi and Yang(2005)]{radinschi6} Radinschi I. and Yang, I.C. ``On the Energy of String Black Holes,'' (2005) New Developments in String Theory Research, ed. Susan A. Grece, New York: Nova Science.
  \bibitem[Riess, et. al.(1998)]{riess} Riess, A.G., et al. [Supernova Search Team Collaboration] (1998)  Astron. J. 116, 1009.
\bibitem[Sharif and Jawad(2011)]{sharif} Sharif, M. and Jawad, A. (2011) Astrophys. and Space Sci. 331, 1, 321.
\bibitem[Vagenas(2006)]{vagenas} Vagenas, E.C. (2006) Mod. Phys. Lett. A21, 1947.
\bibitem[Virbhadra(1990)]{virbhadra1} Virbhadra, K.S. (1990) Phys. Rev. D41, 1086.
\bibitem[Virbhadra(1999)] {virbhadra2} Virbhadra, K.S. (1999) Phys. Rev. D60, 104041.
\bibitem[(1972)] {weinberg} Weinberg, S. (1972) Gravitation and Cosmology: Principles and Applications of the General Theory of Relativity, Wiley, New York, 165.
 \bibitem[(2011)] {Yazadjiev} Yazadjiev, S. (2011) Phys. Rev. D. 83, 127501.
   
    \end{thebibliography}
\end{document}